\documentclass{tlp}

\usepackage{amsmath}
\usepackage{graphicx}
\usepackage{multirow}

\usepackage{subcaption}
\usepackage{amsfonts}
\usepackage{amsthm}
\usepackage{hyperref}
\usepackage[dvipsnames]{xcolor}

\newtheorem{theorem}{Theorem}

\newtheorem{definition}{Definition}
\theoremstyle{definition}
\newtheorem{example}{Example}
\theoremstyle{remark}

\newtheorem*{note*}{Note}
\newtheorem{remark}{Remark}
\newtheorem*{remark*}{Remark}

\begin{document}

\lefttitle{J. Franc\`{e}s de Mas and J. Bowles}

\jnlPage{1}{19}
\jnlDoiYr{2024}
\doival{10.1017/xxxxx}

\title[A novel framework for systematic PL formula simplification based on EGs]{A novel framework for systematic propositional formula simplification based on existential graphs}

\begin{authgrp}
\author{\gn{Jordina} \sn{Franc\`{e}s de Mas}}
\affiliation{School of Computer Science, University of St Andrews, Scotland, UK\\ \email{jfdm2@st-andrews.ac.uk}}
\author{\gn{Juliana} \sn{Bowles}\thanks{Bowles is partially supported by Austrian FWF Meitner Fellowship M-3338 N.}}
\affiliation{School of Computer Science, University of St Andrews, Scotland, UK\\ \email{jkfb@st-andrews.ac.uk}}
\affiliation{Software Competence Centre Hagenberg (SCCH), Austria}
\end{authgrp}

\history{\sub{11 02 2024;} \rev{22 05 2024;} \acc{xx xx xxxx}}

\maketitle

\begin{abstract}
    This paper presents a novel simplification calculus for propositional logic derived from Peirce's existential graphs' rules of inference and implication graphs. Our rules can be applied to propositional logic formulae in nested form, are equivalence-preserving, guarantee a monotonically decreasing number of variables, clauses and literals, and maximise the preservation of structural problem information.
    Our techniques can also be seen as higher-level SAT preprocessing, and we show how one of our rules (TWSR) generalises and streamlines most of the known equivalence-preserving SAT preprocessing methods. In addition, we propose a simplification procedure based on the systematic application of two of our rules (EPR and TWSR) which is solver-agnostic and can be used to simplify large Boolean satisfiability problems and propositional formulae in arbitrary form, and we provide a formal analysis of its algorithmic complexity in terms of space and time. Finally, we show how our rules can be further extended with a novel n-ary implication graph to capture all known equivalence-preserving preprocessing procedures.\vspace{5pt}
    
    \noindent\textit{Under consideration in Theory and Practice of Logic Programming (TPLP).}
\end{abstract}

\begin{keywords}
Equivalence-preserving preprocessing, Propositional logic, Existential graphs, Knowledge representation
\end{keywords}

\section{Introduction}
\label{sec:intro}
Propositional logic simplification is closely related to the reduction of complex Boolean algebraic expressions, logic circuits' minimisation, and Boolean satisfiability (SAT) preprocessing techniques. Simplification is crucial to reduce the complexity of a problem, which makes it easier to understand, reason about, and work with. Minimising the size of a problem also reduces memory usage and speeds up solving times~\citep{DBLP:conf/sat/SorenssonB09}, as fewer steps are required to reach a solution or proof. Unfortunately, existing minimisation algorithms for Boolean expressions such as Karnaugh maps~\citep{karnaugh1953map} or the Quine–McCluskey algorithm~\citep{mccluskey1956minimization} become exponentially inefficient as the number of variables grows, making them unusable for large formulas. To minimise bigger Boolean problems, we can only resort to suboptimal heuristic methods, such as those used by the ESPRESSO logic minimiser~\citep{rudell1989logic} or the state-of-the-art logic synthesis and verification tool ABC~\citep{DBLP:conf/cav/BraytonM10}. Intuitively, all equivalence-preserving simplifications can also be achieved by applying the axioms of Boolean algebra, known logical equivalences and inference rules (such as those in sequent calculus and natural deduction) in nontrivial ways. However, there are currently no known generic or systematic methods for doing so.

Many simplification techniques in the field of SAT use heuristics too, but a greater emphasis is placed on efficiency so that they can be used on very large problems. 
Preprocessing has in fact become an essential part of SAT solving, with some of the most powerful SAT solvers interleaving preprocessing with solving steps, a technique referred to as \textit{inprocessing}~\citep{DBLP:conf/cade/JarvisaloHB12,DBLP:conf/sat/FazekasBS19}. Despite the already vast body of literature on preprocessing (see Chapter~9 in the latest Handbook of Satisfiability~\citep{DBLP:series/faia/BiereJK21}), there is still much ongoing research into finding efficient rewriting and simplification techniques to expedite or better inform the solving phase (see, e.g.,~\cite{DBLP:conf/sat/Anders22,DBLP:journals/ai/LiXLMLL20,fleury2020cadical,DBLP:conf/ijcai/LuoLXML17}).
All these efforts emphasise the importance and complexity of this topic. However, SAT problems commonly need to be translated into a standard form (usually conjunctive normal form (CNF)) before solving, and most procedures apply to this form only. 
Consequently, most preprocessing techniques in the literature only study or work on CNF formulae. Moreover, this encoding process can be non-equivalence-preserving, result in bigger problems or lead to the loss of important structural properties of the original problem (such as symmetries) which can detrimentally impact the preprocessing and solving phases~\citep{DBLP:conf/sat/EenMS07,DBLP:conf/sat/Anders22}.

As an example, consider the formula $((P\rightarrow Q)\rightarrow P)\rightarrow P$, known as Peirce's Law. Let us reflect on how it can be simplified, proven or disproven. From the extensive set of known logical identities and inference rules, which ones do we need to use in this case and in what order? After studying logic, memorising all the rules and techniques, and practicing proving logic statements like this, one naturally
gains experience and intuition on how to detect patterns, common expressions or use clever tactics to find a proof more easily, as is done in mathematical reasoning. A brute force approach of just applying all rules in all possible ways is clearly not feasible as the problem size grows, and can even lead to non-termination, given that some rules have the opposite effect of others, meaning that we could get stuck in an infinite loop. When logicians prove theorems, they usually look ahead to the desired conclusion and try to guide the proof in that direction, but it is not necessarily clear in advance what problem transformations will make it easier, or conversely harder, to find a proof. For example, applying the distributive law increases the size of the problem by certain measures, but can facilitate the application of 
rules which cannot be applied in the factorised form.

\begin{figure}[h]
    \centering
    \includegraphics[width=0.5\linewidth]{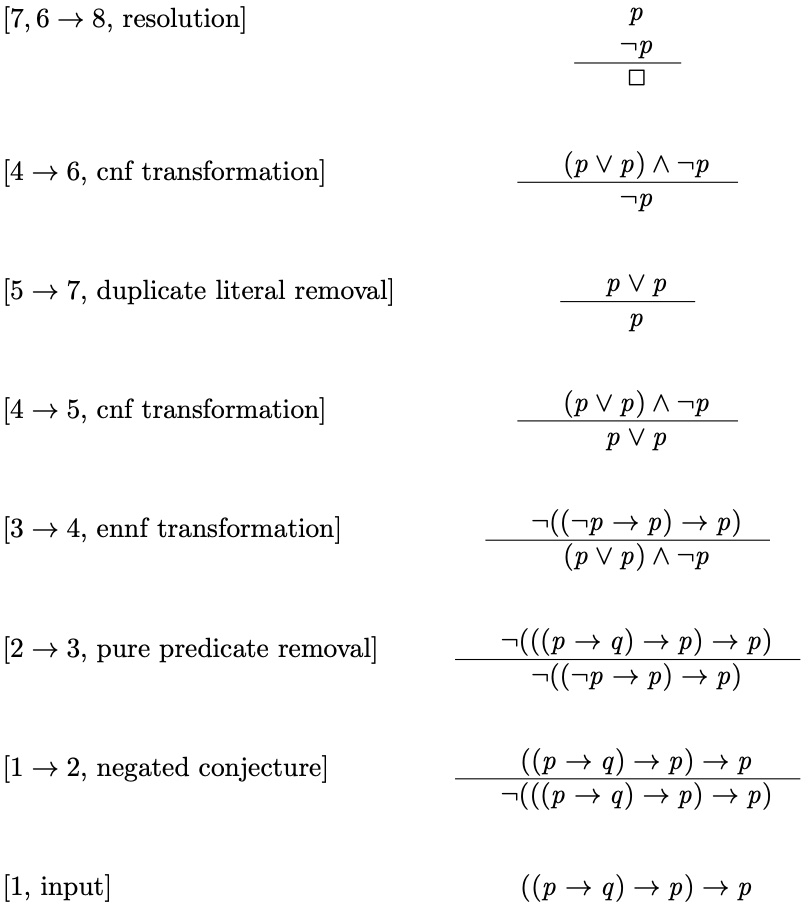}
    \caption{Vampire proof of Peirce’s Law, extracted from~\cite{DBLP:conf/cade/Reger16}. The proof starts from the bottom, and the text within square brackets indicates the step order as well as the name of the transformation or inference rule applied.}
    \label{fig:vampi}
\end{figure}

Nowadays, proof assistants and theorem provers can help us to find or check a formal proof. Even though expertise and insight are still very much required ---in addition to the added non-trivial software proficiency--- these tools can make it easy to find a proof to simple problems like the one suggested above. For example, the automatic theorem prover for first-order classical logic Vampire\footnote{\url{https://vprover.github.io/}} is able to automatically generate a proof by contradiction of Peirce's Law, as shown in Fig.~\ref{fig:vampi}. Note, however, that it requires three heuristic transformations to normal forms ---namely conjunctive normal form (CNF) and equivalence negation normal form (ENNF)--- to do so.
    
In this paper, we present novel simplification techniques for zeroth-order logic derived from Peirce's existential graphs' rules of inference and implication graphs. Our rules can also be seen as SAT equivalence-preserving preprocessing techniques applicable to arbitrary propositional logic formulae (not only in CNF) that guarantee a monotonically decreasing number of variables, clauses and literals and maximise the preservation of structural problem information. Existential graphs offer a fresh view of preprocessing never explored before that allowed us to independently rediscover many simplification techniques known in the world of SAT preprocessing, understand the underlying connections between apparently distinct methods, and generalise many of them. We compare our rules with the state-of-the-art and discuss their advantages and limitations. In particular, our last rule (TWSR) can efficiently emulate complex combinations of preprocessing techniques, and we propose even more powerful extensions to it.

The remainder of our paper is structured as follows. Section~\ref{sec:bk} introduces basic concepts and notation on existential graphs, propositional logic and implication graphs that will be used in the rest of the paper. In Section~\ref{sec:calc}, we present our simplification rules and explain their properties. In Section~\ref{sec:fw}, we discuss in detail the algorithmic complexity of our approach and point to future work, including a generalisation of our TWSR rule. Section~\ref{sec:conc} ends the paper with some concluding remarks.

\section{Background and notation}
\label{sec:bk}
\begin{figure}[h]
     \centering
     \begin{subfigure}[t]{0.45\textwidth}
         \centering
         \includegraphics[height=0.75cm]{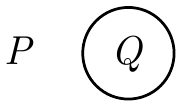}
         \caption{EG of $P \land \neg Q$ or, equivalently, $\neg Q \land P$.
         }
         \label{fig:EGa}
     \end{subfigure}
     \hfill
     \begin{subfigure}[t]{0.45\textwidth}
         \centering
         \includegraphics[height=1.5cm]{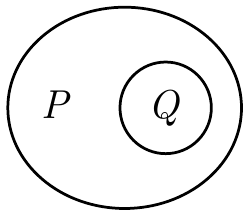}
         \caption{EG of $\neg(P \land \neg Q)$ or, equivalently, $\neg P \lor Q$, or $P\rightarrow Q$.}
         \label{fig:EGb}
     \end{subfigure}
     \begin{subfigure}[t]{0.45\textwidth}
        \centering
        \includegraphics[height=1.75cm]{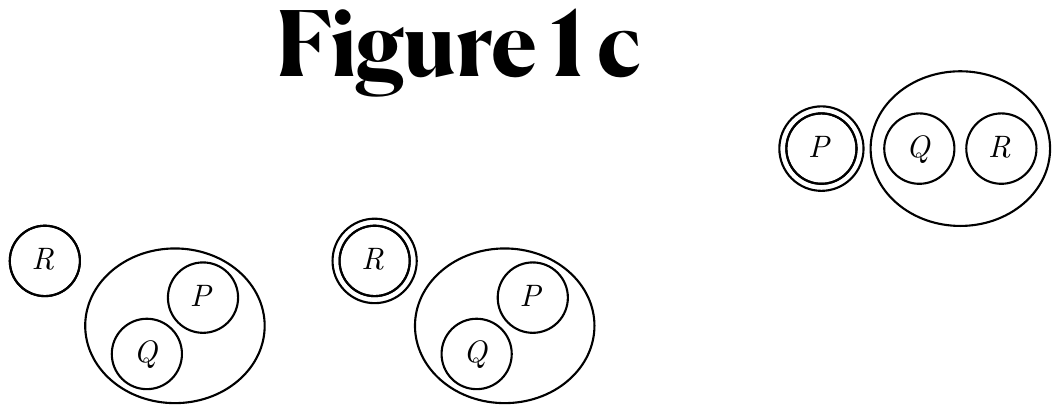}
        \caption{EG of $\neg R \land \neg(\neg Q \land \neg P)$ or, equivalently, $(P\lor Q)\land\neg R$, or $\neg R\land (\neg P\rightarrow Q)$, or $\neg(R\lor(\neg P\land\neg Q))$.}
        \label{fig:EGc}
     \end{subfigure}
     \hfill
     \begin{subfigure}[t]{0.45\textwidth}
        \centering
        \includegraphics[height=1.5cm]{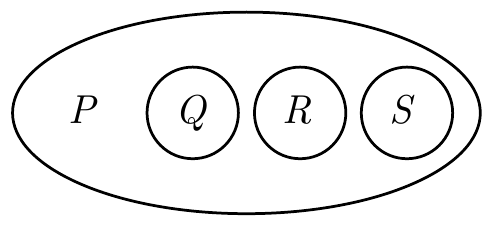}
        \caption{EG of $\neg(P \land \neg Q \land \neg R \land \neg S)$, which can be linearly represented by many equivalent formulations, such as $P \rightarrow (Q \lor R \lor S)$, or $(P \land \neg Q) \rightarrow (R \lor  S)$, or $(P \land \neg Q \land \neg R) \rightarrow S$, or $\neg P \lor Q \lor R \lor S$.}
        \label{fig:EGd}
     \end{subfigure}
    \caption{Examples of existential graphs representing different propositional logic formulae.}
    \label{fig:EGintro}
\end{figure}

Modern formal logics, including formulations of SAT problems, are usually presented in a symbolic and linear fashion. \textbf{Existential graphs} (EGs)~\citep{sowa2011peirce,DBLP:journals/jolli/Shin99}, by contrast, are a \textit{non-symbolic} and \textit{non-linear} system for logical expressions, where propositions are drawn on a two-dimensional sheet, negation is represented by oval lines (aka \textit{cuts}), and all elements contained in the same area (delimited by negation lines) are in implicit conjunction.
We say that an area is \textit{evenly} (resp. \textit{oddly}) \textit{nested} if it is enclosed by an even (resp. odd) number of cuts. Note that a blank sheet of assertion denotes \textit{true} ($\top$), has nesting level $0$ and therefore it is assumed to be evenly enclosed.
In Fig.~\ref{fig:EGintro}, we present a few introductory examples illustrating that the combination of these simple primitives suffices to express any propositional logic formula.\vspace{-7pt}

EGs cannot only represent propositional logic formulae, but they are, in fact, equivalent to sentential languages\footnote{Extensions of EGs allow for the representation of first-order logic and higher-order logic formulas, but these are beyond the scope of this paper (for more details, see e.g.~\cite{dau2008mathematical,roberts1973existential}).}~\citep{zeman1964graphical,roberts1973existential}. Even if this diagrammatic notation is less commonly used, it offers a simple, elegant, and easy-to-learn alternative that comes with a sound and complete deductive system which has many advantages over the traditional Gentzen's rules of \textit{natural deduction} and \textit{sequent calculus} and, thus, over all other rewriting methods based on these. Most significantly, EGs' inference rules are symmetric, and no bookkeeping of previous steps is required in the process of inferring a graph from another. Furthermore, as illustrated in Fig.~\ref{fig:EGintro}, EGs' graphical non-ordered nature provides a canonical representation of a logical formula that, in linear notation, can take many equivalent forms. Another key and perhaps underexplored advantage is that the visual representation allows for the recognition of patterns that would otherwise be obscured in nested clausal form. Finally, since EGs are easily transferred to any notation, they can help clarify the relationships between diverse reasoning methods, such as resolution and natural deduction and, ultimately, help understand the foundations of proof theory.

We assume the reader is familiar with basic notions of \textit{propositional logic} and the \textit{Boolean satisfiability problem} (\textbf{SAT}). In what follows, we will use a subscript to indicate the size of a clause (e.g. $C_8$ refers to a clause of size $8$) and, given a clause $C_n=(l_1\lor l_2\lor\cdots\lor l_n)$, we refer to the set of its literals as $\mathrm{lit}(C_n)=\{l_1,l_2,\cdots,l_n\}$.

\begin{figure}[h]
    \centering
    \includegraphics[width=0.5\textwidth]{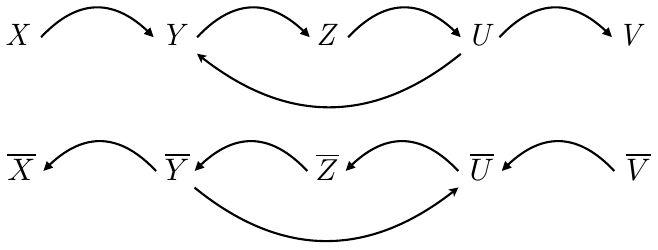}
    \caption{BIG of $\varphi=(\overline{X}\lor Y)\land(\overline{Y}\lor Z)\land(\overline{Z}\lor U)\land(\overline{U}\lor V)\land(\overline{U}\lor Y)$.}
    \label{fig:big}
\end{figure}

As shown in Fig.~\ref{fig:EGintro}, nested cuts can be interpreted as implication, so it is easy to see that every binary clause, e.g. $C_2=(x\lor y)$, admits the following two interpretations: $(\neg x\rightarrow y)$ and its logically equivalent contrapositive $(\neg y\rightarrow x)$. Thus, binary clauses provide the information needed to build what is known as the \textbf{binary implication graph} (BIG)~\citep{DBLP:journals/ipl/AspvallPT79} of the formula, which is a directed graph where edges represent implication, and nodes are all the literals occurring in binary clauses and their negations (see Fig.~\ref{fig:big}).

If there exists a directed path from vertex $x$ to vertex $y$, we say that $y$ is a \textbf{descendant} of (or \textit{reachable} from) $x$, and we say that $x$ is an \textbf{ancestor} of $y$. We will denote the set of all descendants (resp. ancestors) of a vertex $z$ by $\mathrm{des}(z)$ (resp. $\mathrm{anc}(z)$).
We say that two vertices $u$ and $v$ in a directed graph $G$ are strongly connected iff
there are directed paths from $u$ to $v$ and from $v$ to $u$.
A subgraph $S\subseteq G$ is a \textbf{strongly connected component} if all its vertices are strongly connected and they are not strongly connected with any other vertex of $G\setminus S$ (i.e. every vertex in $S$ is reachable from every other vertex in $S$ and $S$ is maximal with this property wrt $G$). Every strongly connected component corresponds to an \textbf{equivalence class}. Two elements belong to the same equivalence class iff they are equivalent. We can choose one of the elements in the class as a \textbf{class representative} to refer to the whole equivalence class.

\section{Preprocessing framework}
\label{sec:calc}

\subsection{Simplification rules}\label{sr-subsec}
Charles S. Peirce (1839-1914) provided the following sound and complete set of inference rules for EGs (for more details, see e.g.~\cite{peirce1909existential,sowa2011peirce,DBLP:journals/jolli/Shin99,dau2008mathematical}), where an \textit{EG-element} is any arbitrary clause\footnote{\label{note:clause}We refer here to the generic notion of `clause', understood as a propositional formula consisting of a finite collection of literals and logical connectives.} expressed as an EG:
\begin{enumerate}
    \item \begin{enumerate}
        \setcounter{enumii}{8}
        \item \begin{description}
            \item[Insertion:] in an odd area (nesting level $2k+1, k\in\mathbb{N}$), we can draw any EG-element.
            \end{description}
        \setcounter{enumii}{4}
        \item \begin{description}
            \item[Erasure:] any EG-element in an even area (nesting level $2k, k\in\mathbb{N}$) can be deleted.
            \end{description}
    \end{enumerate}
    \item \begin{enumerate}
        \setcounter{enumii}{8}
        \item \begin{description}
            \item[Iteration:] any EG-element in an area $a$ can be duplicated in $a$ or in any nested areas within $a$.
            \end{description}
        \setcounter{enumii}{4}
        \item \begin{description}
            \item[Deiteration:] any EG-element whose occurrence could be the result of iteration may be erased.
            \end{description}
    \end{enumerate}
    \item \begin{enumerate}
        \setcounter{enumii}{8}
        \item \begin{description}
            \item[Double Cut Insertion:] a double negation may be drawn around any collection of zero or more EG-elements in any area.
            \end{description}
        \setcounter{enumii}{4}
        \item \begin{description}
            \item[Double Cut Erasure:] any double negations can be erased.
            \end{description}
    \end{enumerate}
\end{enumerate}
We investigate the reversibility properties of Peirce's EGs inference rules and their combinations in order to determine which sequences of rule applications and underlying restrictions ensure non-increasing model equivalence. Note that rules $2i$ \& $2e$ and $3i$ \& $3e$ are mutually reversible and, thus, preserve equivalence, whilst the first two rules are not.
To the best of our knowledge, the EG calculus has only been studied as a proof system and used to prove logical assertions by inference \citep{dau2008mathematical,sowa2011peirce}, but it has never been used as a simplification method nor restricted to equivalence-preserving rules.

In what follows, we present the definitions of each of our rules alongside a visual exemplification, and compare them to existing preprocessing techniques. Throughout, let $\varphi$ be an arbitrary propositional formula and BIG($\varphi$) denote its binary implication graph. 

\subsubsection{Singleton wipe}

This rule is equivalent to a combination of Shin's \textit{rules 1} and \textit{2}~\citep{DBLP:journals/jolli/Shin99}, and to Peirce's EGs \textit{deiteration rule} ($2e$ above) restricted to single EG-elements (unit clauses) and extended to account for the generation of empty cuts. Most importantly, it can be seen as a generalisation of \textit{unit propagation} applicable over arbitrary formulae, not only in CNF.
\begin{definition}[Singleton wipe rule (SWR)]\label{def:swr}
    Any copy of a single EG-element (i.e. a literal) present in the same or any nested (inner) areas can be erased, and, if an empty cut is generated, the area containing the empty negation oval shall be deleted.
\end{definition}
Note that, in linear notation, clausal levels can be distinguished by parenthesis (either explicit or implicit in accordance with the precedence of logical operators). Moreover, note that empty cuts are generated when the negation of the propagated singleton is encountered.

\begin{example}\label{ex:sw}
    Let $\varphi = P \wedge ((A \wedge D \wedge P \wedge (A \rightarrow B) \wedge (\neg C \vee D)) \vee (P \wedge Q \wedge R) \vee T \vee (S \wedge T) \vee \neg(X \rightarrow (X \wedge Y \wedge Z))) \wedge \neg T$. When trying to simplify $\varphi$, several singletons can be propagated in the same or inner levels, namely $A, D, P, \neg T,$ and $X$. This is not easy for a human reader to detect if the formula is expressed in linear form, but it becomes apparent when expressed as an EG (see Fig.~\ref{fig:SWR}). If we apply SWR to $\varphi$, we obtain the following simplified equivalent formula: $\varphi'=\mathrm{SWR}(\varphi)=P \wedge \neg T \wedge ((A \wedge B \wedge D) \vee (Q \wedge R) \vee (X \wedge (\neg Y \vee \neg Z)))$. Note that transforming $\varphi$ to CNF (with trivial tautology removal) would result in a formula of $30$ clauses of sizes $1$ to $5$ with a total of $124$ literals and $11$ variables, where traditional unit propagation could only be applied to the unit clauses $P$ and $\neg T$, and the resulting (partially) simplified formula would have $26$ clauses of sizes $1$ to $5$ with a total of $98$ literals and $11$ variables. Instead, if we transform $\varphi'$ to CNF, we obtain a formula with $10$ variables, $14$ clauses of sizes $1$ to $4$, and a total of $44$ literals.
    
    \begin{figure}[h]
        \centering
        \includegraphics[width=0.75\textwidth]{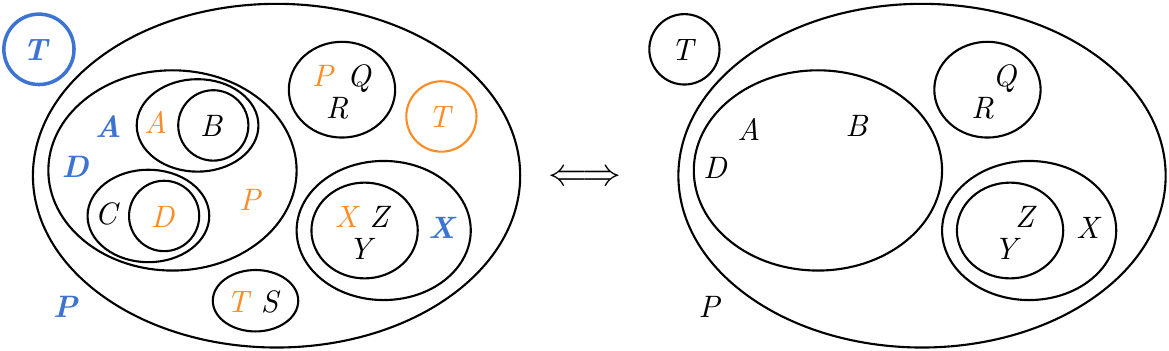}
        \caption{EG of $\varphi$ (left), where the EG-elements coloured in orange can be simplified by propagating their outermost instances, marked in blue, to obtain $\varphi'$ (right).}
        \label{fig:SWR}
    \end{figure}
    
\end{example}

\begin{remark}
SWR effectively tells us that each EG-element (literal, clause, or subclause) contributes to the truth value of the whole EG (i.e. the whole formula) at its outermost occurrence, and the presence or absence of more deeply nested instances is irrelevant.
\end{remark}

\subsubsection{Equivalence projection}
This rule can be achieved by nontrivial applications of the iteration, deiteration and cut rules ($2i$, $2e$, $3i$ and $3e$ from Section~\ref{sr-subsec}) and noticing that two propositional variables $x$ and $y$ are equivalent (resp. opposites) whenever we have the following two clauses: $(x\lor\overline{y})\land(\overline{x}\lor y)$ (resp. $(x\lor y)\land(\overline{x}\lor\overline{y})$). This is obvious from the implication interpretation of these clauses as EGs that we saw in Section~\ref{sec:bk}.

Trying all possible (de)iterations of binary clauses within each other is clearly impractical, and requires a nontrivial search for candidate EG-elements and backtracking. However, if we use the BIG of the formula to inform our procedure, it becomes straightforward and efficient.

\begin{definition}[Equivalence projection rule (EPR)]
     Let $\phi$ be an arbitrary subformula of $\varphi$. Every \emph{strongly connected component} of $\mathrm{BIG(\phi)}$ corresponds to an \emph{equivalence class}, and all literals in the equivalence class in the same or greater nesting level within $\phi$ can be substituted by their \emph{class representative}. If a cut contains multiple elements of the equivalence class, they can all be substituted by a single instance of the representative literal. If a cut contains both an element of the equivalence class and the negation of an element in the equivalence class, it can be deleted. The $\mathrm{BIG(EPR(\phi))}$ of the remaining binary clauses is guaranteed to be acyclic and corresponds to the \emph{condensation} of $\mathrm{BIG(\phi)}$.
\end{definition}
The deletion of literals and clauses in EPR can be seen as a nested application of the SWR rule (see Definition~\ref{def:swr}) performed immediately after a substitution step. 
The replacement part of EPR is in fact equivalent to a well-known preprocessing technique called \textit{equivalent-literal substitution} (ELS)~\citep{DBLP:conf/dimacs/GelderT93,DBLP:conf/aaai/Li00}, but ours does not require a formula to be in CNF. Moreover, our rule is equivalence-preserving (unlike ELS) since we keep the information of any equivalence classes, and both substitution and simplification are applied in one step. Additionally, the BIG built and updated during the application of this rule will inform other preprocessing techniques, so the effort of building the graph will be reused and further exploited.

\paragraph{Nested equivalence projection}
Equivalence projection can be applied not only at the formula level but also within nested cuts, which enhances its reduction powers and makes EPR applicable to formulae in forms other than CNF too (see Example~\ref{ex:nep1}). To do so, we maintain local implication graphs corresponding to the binary clauses present at each nesting level.
\begin{example}\label{ex:nep1}
    Let $\varphi=(\overline{X}\lor\overline{Y}\lor\overline{A})\land\neg((A\rightarrow B)\land(B\rightarrow A)\land(\overline{X}\lor\overline{Y}\lor\overline{B}))$. We generate the BIG of any area with binary clauses, which in this case is only the big level-$1$ area. The nested BIG contains a strongly connected component $[A]=\{A,B\}$, so the innermost $B$ can be substituted by the representative $A$. The resulting ternary clause can then be wiped by the deiteration rule (and our rules to come), leading to the negation of the equivalence found (see Fig.~\ref{fig:NEP1}). Hence $\varphi$ can be simplified to $\varphi'=\mathrm{EPR}(\varphi)=(\overline{X}\lor\overline{Y}\lor\overline{A})$ with $[A]=\{A,\overline{B}\}$.
    
    \begin{figure}[h]
        \centering
        \includegraphics[width=0.75\textwidth]{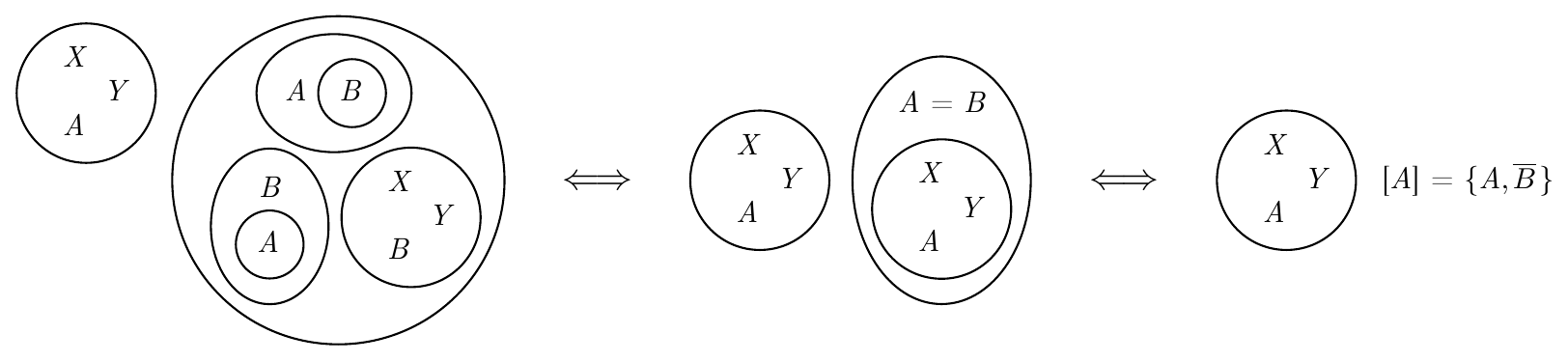}
        \caption{EGs of $\varphi=(\overline{X}\lor\overline{Y}\lor\overline{A})\land\neg((A\rightarrow B)\land(B\rightarrow A)\land(\overline{X}\lor\overline{Y}\lor\overline{B}))$ and its reduction to $\varphi'=(\overline{X}\lor\overline{Y}\lor\overline{A})$ with $[A]=\{A,\overline{B}\}$ after nestedly applying EPR and then deiteration.}
        \label{fig:NEP1}
    \end{figure}
    
\end{example}

Even higher reductions can be achieved if we consider equivalence ---and, as we will see, implication chains--- in the union of nested implication graphs (see Example~\ref{ex:nep2}).
\begin{example}\label{ex:nep2}
    Let $\varphi=(\overline{X}\lor\overline{Y}\lor\overline{A})\land(A\rightarrow B)\land(B\rightarrow A)\land\neg((C\rightarrow B)\land(B\rightarrow C)\land(\overline{X}\lor\overline{Y}\lor\overline{C}))$. We generate the BIG of any area with binary clauses, which in this case are the outermost area and the biggest level-$1$ area. The outer BIG contains a strongly connected component $[A]=\{A,B\}$, and the nested BIG contains a strongly connected component $[B]=\{B,C\}$. The union of both BIGs results in the strongly connected component $[A]=\{A,B,C\}$, so the inner ternary clause can be wiped by a combination of EPR and the deiteration rule. Thus, $\varphi$ can be simplified to $\varphi'=\mathrm{EPR}(\varphi)=(\overline{X}\lor\overline{Y}\lor\overline{A})$ with $[A]=\{A,B,\overline{C}\}$ (see Fig.~\ref{fig:NEP2}).
    \begin{figure}[h]
        \centering
        \includegraphics[width=\textwidth]{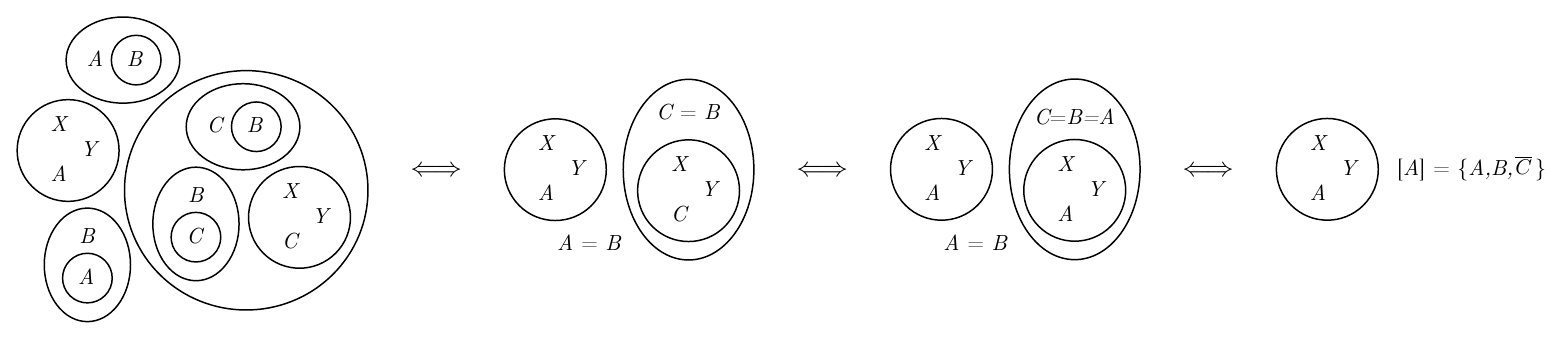}
        \caption{EGs of $\varphi=(\overline{X}\lor\overline{Y}\lor\overline{A})\land(A\rightarrow B)\land(B\rightarrow A)\land\neg((C\rightarrow B)\land(B\rightarrow C)\land(\overline{X}\lor\overline{Y}\lor\overline{C}))$ and its reduction to $\varphi'=(\overline{X}\lor\overline{Y}\lor\overline{A})$ with $[A]=\{A,B,\overline{C}\}$ after nestedly applying EPR to the union of nested BIGs and then deiteration.}
        \label{fig:NEP2}
    \end{figure}
\end{example}

\subsubsection{Transitive reduction}\label{ssec:tr}
After applying EPR until completion, the BIG is guaranteed to be acyclic since all equivalences (and so strongly connected components) have been condensed into a representative literal (or node). However, the formula can still contain redundant binary clauses.
In order to remove those, we compute the \textit{transitive reduction} (TR) of the BIG, which coincides with its minimum equivalent graph.
Even if TR is a well-known concept, the EGs viewpoint allows us to apply this technique in a nested form and be able to detect transitive redundancies in arbitrary formulas (see Example~\ref{ex:ntr}). We refer to this systematic application of TR across nesting levels and BIG unions as TRR. As in the EPR case, the same results can be achieved by nontrivial applications of the iteration, deiteration and cut rules, but these are only efficient if guided by the BIG.
\begin{example}\label{ex:ntr}
    Let $\varphi=(\overline{A}\lor B)\land(\overline{B}\lor C)\land\neg((C\lor\overline{A})\land X\land Y)$. Then the level-0 BIG contains the implication chain $A\implies B\implies C$, and the level-1 binary clause is equivalent to $A\rightarrow C$. Computing the TR of the union of both BIGs shows that the inner binary clause is redundant and can be deleted. As illustrated in Fig.~\ref{fig:NTR}, we then obtain the equivalent simplified formula $\varphi'=(\overline{A}\lor B)\land(\overline{B}\lor C)\land(\overline{X}\lor\overline{Y})$.
    \begin{figure}[h]
        \centering
        \includegraphics[width=0.5\textwidth]{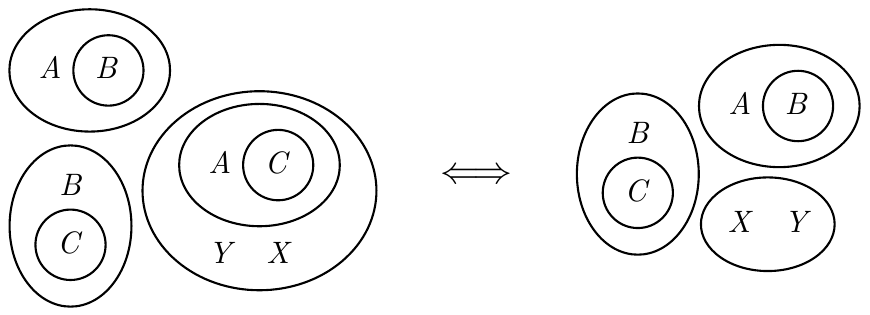}
        \caption{EGs of $\varphi=(\overline{A}\lor B)\land(\overline{B}\lor C)\land\neg((C\lor\overline{A})\land X\land Y)$ (left), and its equivalent reduction $\varphi'=\mathrm{TRR}(\varphi)=(\overline{A}\lor B)\land(\overline{B}\lor C)\land(\overline{X}\lor\overline{Y})$ (right).}
        \label{fig:NTR}
    \end{figure}
\end{example}

\subsubsection{Opposite singletons implication}\label{ssec:osir}
\begin{definition}[Opposite singletons implication rule (OSIR)]
    If a directed path in \emph{BIG}$(\varphi)$ contains a literal $l$ and later its negation $\overline{l}$, then all the literals including and after the consequent ($\overline{l}$) evaluate to true, and all the literals in the implication path before and including the antecedent ($l$) evaluate to false. The Boolean variables ``collapsing'' in opposite chains will be equal by construction, so only one side of the implication path needs to be evaluated. All evaluated literals can be added as singletons and propagated accordingly by the \emph{SWR} rule.
\end{definition}
Note that this preprocessing step can also be seen as a backtrack-free (i.e. non-look-ahead) exhaustive version of \textit{failed-literal probing} (FLP)~\citep{DBLP:journals/endm/Berre01} over all (implicit and explicit) binary clauses, where we do not need to test candidate literals that might not lead to any new knowledge after a whole round of unit propagation. A strategy to make all possible unit-clause inferences from the BIG was proposed in~\cite{DBLP:journals/amai/Gelder05}, but their approach may add redundant, unnecessary clauses to the formula. As the previous rules, OSIR can also be applied to arbitrary nesting levels, so it does not require a formula to be in CNF (see Example~\ref{ex:osir}).
\begin{example}\label{ex:osir}
    Let $\varphi=(\overline{X}\lor\overline{Z})\land((B\land\overline{C})\lor(X\land Y\land\overline{A})\lor(A\land\overline{B})\lor(P\land A\land Q)\lor(C\land A))$. The BIG of the $1$-nested biggest subformula contains the following implication chain: $A\implies B\implies C\implies\overline{A}$, so we can apply the OSIR to it and derive $\overline{A}$, which can be added as a nested singleton and later propagated using the SWR rule (see Fig.~\ref{fig:OSIR} below) to obtain the much simpler formula $\varphi'=\mathrm{OSIR}(\varphi)=(\overline{X}\lor\overline{Z})\land((B\land\overline{C})\lor(X\land Y)\lor A)$.
    
    \begin{figure}[h]
        \centering
        \includegraphics[width=\textwidth]{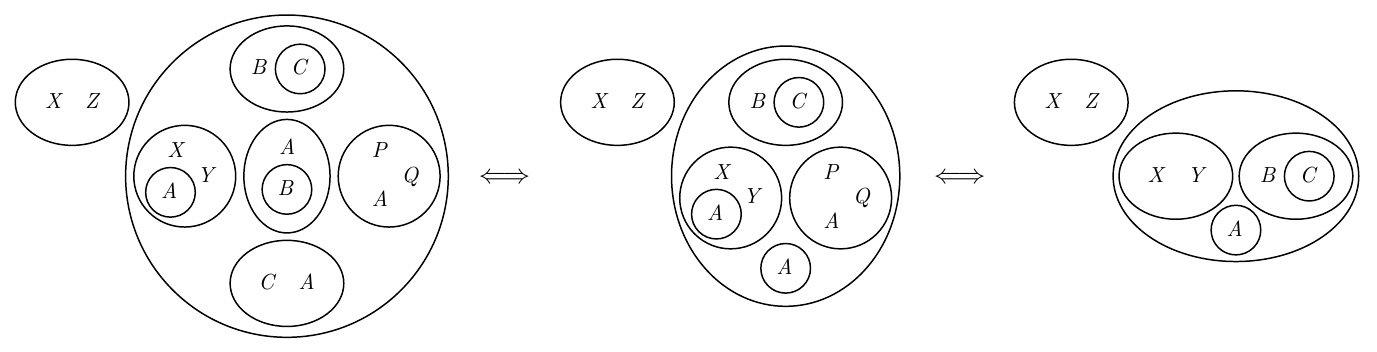}
        \caption{EGs showing the application of OSIR to $\varphi=(\overline{X}\lor\overline{Z})\land((B\land\overline{C})\lor(X\land Y\land\overline{A})\lor(A\land\overline{B})\lor(P\land A\land Q)\lor(C\land A))$ and the subsequent SWR application where the new singleton $\overline{A}$ is propagated to obtain the equivalent reduced formula $\varphi'=(\overline{X}\lor\overline{Z})\land((B\land\overline{C})\lor(X\land Y)\lor A)$.}
        \label{fig:OSIR}
    \end{figure}
    
\end{example}

\subsubsection{Tuple wipe and subflip}\label{ssec:twsr}
After understanding the basics of Peirce's EGs rules and binary implication graphs, we can now introduce a rule which, in fact, generalises all the previous rules (except the non-reductive part of EPR).
\begin{definition}[Tuple wipe and subflip rule (TWSR)]
    Let $C_n,D_m\in\varphi$ be two (sub)clauses of size $n\leq m$. Let $c$ be the nesting level of $C_n$, and $D_m$ be either in the same area as $C_n$ or in a nested area within $c$. Let $\mathrm{des}(l)$ be the set of descendants of a literal $l$ in $\mathrm{BIG}(\varphi)$. Let $\mathrm{lit}(C_n)=\{p_1,\dots,p_n\}$ and $\mathrm{lit}(D_m)=\{q_1,\dots,q_m\}$.
    
    \emph{(1)} If for each $i\in\{1,...,n-1\}$ either $p_i=q_i$ or $q_i\in \mathrm{des}(p_i)$, and $p_n=q_n$ or $q_n\in \mathrm{des}(p_n)$, then $D_m$ can be deleted from $\varphi$.
    
    \emph{(2)} If for each $i\in\{1,...,n-1\}$ either $p_i=q_i$ or $q_i\in \mathrm{des}(p_i)$, and $p_n=\overline{q_n}$ or $\overline{q_n}\in\mathrm{des}(p_n)$, then $q_n$ can be deleted from $D_m$. 
\end{definition}
Note that we need to specify the cases $p_n=q_n$ and $p_n=\overline{q_n}$ since it is always the case that $x\rightarrow x$ and $\overline{x} \rightarrow\overline{x}$, but these tautologies are not added to the BIG in order to keep it redundancy- and tautology-free.

As before, our rule can be applied to arbitrary nesting levels, and so it does not require a formula to be in CNF (see Example~\ref{ex:twsr}).
\begin{example}\label{ex:twsr}
    Let $\varphi=(A\rightarrow E)\land(B\rightarrow F)\land(C\rightarrow G)\land\neg((\overline{A}\lor \overline{B}\lor G\lor \overline{H})\land(\overline{A}\lor \overline{B}\lor \overline{C}\lor \overline{D})\land(\overline{E}\lor \overline{F}\lor \overline{G}))$. Note that the BIG of the outermost area applies to the biggest $1$-nested subformula. As illustrated in Fig.~\ref{fig:TWSR}, we can apply TWSR to the clauses of sizes 3 and 4 inside the biggest $1$-nested area and obtain the simplified equivalent formula $\varphi'=(A\rightarrow E)\land(B\rightarrow F)\land(C\rightarrow G)\land((A\land B\land H)\lor(E\land F\land G))$. In particular, the clauses $P_3=(\overline{E}\lor \overline{F}\lor \overline{G})$ and $Q_4=(\overline{A}\lor \overline{B}\lor \overline{C}\lor \overline{D})$ satisfy condition (1) in TWSR's definition, and $P_3$ together with $R_4=(\overline{A}\lor \overline{B}\lor G\lor \overline{H})$ satisfy condition (2).
    \begin{figure}[h]
        \centering
        \includegraphics[width=0.65\textwidth]{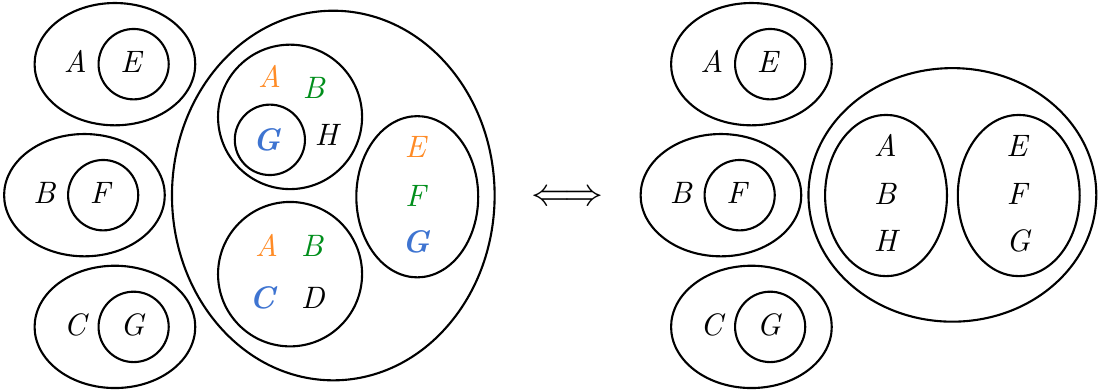}
        \caption{EG of $\varphi=(A\rightarrow E)\land(B\rightarrow F)\land(C\rightarrow G)\land\neg((\overline{A}\lor \overline{B}\lor G\lor \overline{H})\land(\overline{A}\lor \overline{B}\lor \overline{C}\lor \overline{D})\land(\overline{E}\lor \overline{F}\lor \overline{G}))$ (left), where the literals satisfying TWSR's definition conditions are highlighted in matching colours. The equivalent reduced formula resulting from applying TWSR is shown on the right-hand side.}
        \label{fig:TWSR}
    \end{figure}
\end{example}

\begin{figure}[h]
    \centering
    \renewcommand{\thesubfigure}{\roman{subfigure}}
    \begin{subfigure}[t]{0.45\textwidth}
        \centering
        \includegraphics[width=\textwidth]{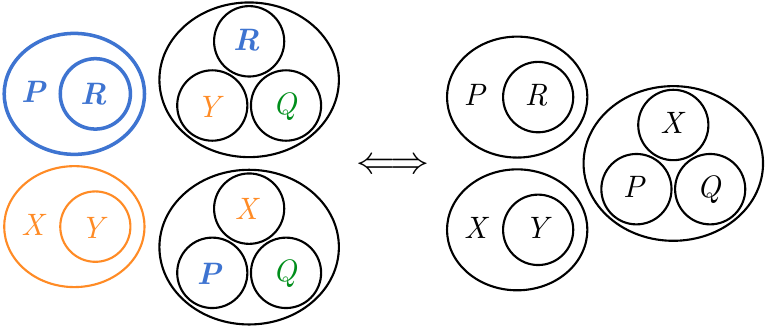}
        \caption{Let $\varphi=(P\rightarrow R)\land(X\rightarrow Y)\land(X\lor Q\lor P)\land(R\lor Y\lor Q)$. Let $C_3=(X\lor Q\lor P)$ be the penultimate clause and $D_3=(R\lor Y\lor Q)$ be the last clause in $\varphi$. These satisfy condition (1) in TWSR's definition, so $D_3$ can be deleted.}
        \label{fig:OFLR1}
    \end{subfigure}
    \hfill
    \begin{subfigure}[t]{0.45\textwidth}
        \centering
        \includegraphics[width=\textwidth]{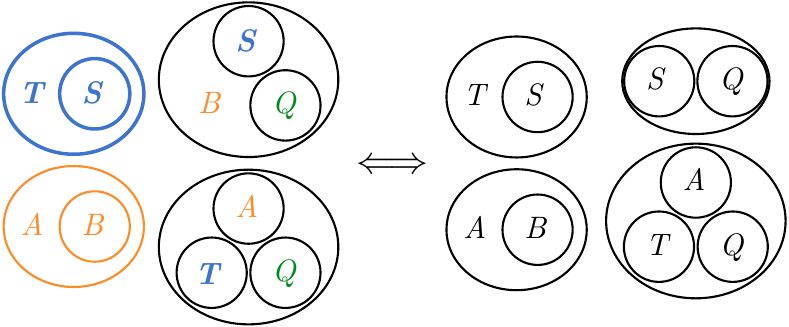}
        \caption{Let $\varphi=(T\rightarrow S)\land(A\rightarrow B)\land(S\lor\overline{B}\lor Q)\land(A\lor Q\lor T)$. Let $C_3=(A\lor Q\lor T)$ and $D_3=(S\lor\overline{B}\lor Q)$. They satisfy condition (2) in TWSR's definition, so the literal $\overline{B}$ can be deleted from $D_3$.}
        \label{fig:OFLR2}
    \end{subfigure}
    \caption{EGs of two applications of the TWSR, where colours highlight related literals (equal ---in green--- or in the same implication chain ---in orange or blue).}
    \label{fig:oflr}
\end{figure}

It is easy to see that TWSR with $n=1$ is equivalent to SWR. TWSR restricted to binary clauses only (i.e. with $n=m=2$) is equivalent to the clause deletion part of EPR and computing the TRR of $\mathrm{BIG}(\varphi)$ when condition (1) is satisfied; and to the literal deletion part of EPR  together with OSIR when condition (2) applies. However, TWSR cannot fully emulate EPR since it cannot perform the non-reductive substitution step (i.e. substitute a literal by its class representative) nor keep equivalence classes information.
Fig.~\ref{fig:oflr} shows two examples of TWSR applications for $n=m=3$ satisfying, respectively, conditions (1) and (2).

Our TWSR is a generalisation of subsumption and self-subsuming resolution since it removes in one pass not only all the explicitly (self)subsumed redundant clauses but also all the \textit{implicitly} (self)subsumed ones. It also generalises unit propagation, transitive reduction and, since it can be applied to clauses of arbitrary sizes, it is strictly more powerful than binary resolution combined with transitive reduction and unit propagation, while guaranteeing a non-increasing number of variables, literals and clauses (i.e. redundant variables, literals or clauses are never added). TWSR can also be seen as a backtrack-free (non-look-ahead, i.e., with no arbitrary assignments) more efficient version of FLP over all literals in the BIG of the formula.

Moreover, TWSR strictly generalises the combination of \textit{hidden subsumption elimination} (HSE)~\citep{DBLP:conf/lpar/HeuleJB10}, \textit{hidden tautology elimination} (HTE)~\citep{DBLP:conf/lpar/HeuleJB10} and \textit{hidden literal elimination} (HLE)~\citep{DBLP:conf/sat/HeuleJB11} in a manner that is guaranteed to never increase the size of the formula. Note that this is not necessarily the case for the aforementioned rules. For example, in order to deduce that clauses are redundant, HTE adds literals instead of removing them and tests if this leads to a tautology. Note in particular that neither of these rules nor their combination could achieve the reduction shown in Example~\ref{ex:twsr}, even if all clauses were at the same nesting level. For formulas in CNF, HTE combined with HSE, computing the transitive closure of BIG($\varphi$) and adding all new binary clauses to $\varphi$ can achieve the same reduction as TWSR with condition (1) only, but clearly at a much higher cost and space complexity.

\subsection{Rules' properties}\label{ss:rp}
Visual proofs using the EG calculus are much easier to follow than symbolic ones, and EGs inference rules have been used to establish the results for the two theorems presented below. However, because of space constraints, we only provide symbolic and `verbal'-EGs skeleton proofs here. Moreover, since the TWSR rule generalises all the others except the substitution part of EPR, we need only provide proofs for these two rules.
\begin{theorem}
    SWR, EPR, TRR, OSIR, and TWSR are all reversible and so equivalence-preserving.
    \begin{proof}[Proof (TWSR)]
        Let $C_n,D_m\in\varphi$ be two (sub)clauses of size $n\leq m$ as in TWSR's definition, with $\mathrm{lit}(C_n)=\{p_1,...,p_n\}$ and $\mathrm{lit}(D_m)=\{q_1,...,q_m\}$. (1) Let $p_i=q_i$ or $q_i\in \mathrm{des}(p_i)$ for each $i\in\{1,...,n\}$. We can iterate $C_n$ inside $D_m$, and all literals in the inner copy satisfying $p_i=q_i$ can be deleted by deiterating their copies in $D_m$. All literals in the inner copy of $C_n$ satisfying $q_i\in \mathrm{des}(p_i)$ can also be deleted as follows: since $q_i\in \mathrm{des}(p_i)$, it means that we have the binary clauses $(\overline{p_i}\lor x_1)\land\dots\land(\overline{x_j}\lor q_i)$, which we can iterate inside $D_m$, and successively deiterate all the consequents until we obtain $D_m\lor p_i\lor x_1\lor\dots\lor x_j\lor\overline{C_n}$. Thus, $D_m$ can be expanded with all $p_i$'s, which can then be deleted from the inner copy of $C_n$, leaving an empty cut which is either in an even or in an odd area. If the empty cut is in an even area, then we have $\bot$ in a conjunction, which evaluates to \texttt{False}. This conjunction is contained in an odd area, and so it can be deleted from the disjunction. If the empty cut is in an odd area, then we have $\top$ in a disjunction, and the whole clause evaluates to \texttt{True}, which can be safely deleted from its implicit surrounding conjunction. (2) Let $p_i=q_i$ or $q_i\in \mathrm{des}(p_i)$ for each $i\in\{1,...,n-1\}$ and $p_n=\overline{q_n}$ or $\overline{q_n}\in \mathrm{des}(p_n)$. We can iterate $C_n$ inside the nesting area of $q_n$ in $D_m$ (with the insertion of a double cut around it if required). If $p_n=\overline{q_n}$, then $\overline{q_n}$ inside the cut can be deiterated from the inner copy of $C_n$, and all the remaining $n-1$ literals can be deleted as in (1). If $\overline{q_n}\in \mathrm{des}(p_n)$, we can deiterate the $n-1$ literals from the inner copy as in (1) and are left with $(\overline{p_n}\lor q_n)$. Since we know that $\overline{q_n}\in \mathrm{des}(p_n)$, we can derive $(\overline{p_n}\lor\overline{q_n})$ from the implication chain as in (1), iterate it inside $q_n$'s cut and deiterate its inner $q_n$ and its inner $p_n$ to obtain an empty cut, which results in deleting $q_n$ from $D_m$, but we then have a $\overline{p}$ in its place. This extra $\overline{p}$ can be deleted by inserting a double cut around it, iterating the $n-1$ implications $q_i\in \mathrm{des}(p_i)$ inside and deiteraiting their $q_1,\dots,q_{n-1}$. This results in $D_m\lor\overline{C_n}$ from which we can deiterate $C_n$ to obtain $D_m\setminus\{q_n\}$. Since we have only used Peirce's iteration ($2i$), deiteration ($2e$) and double cut ($3i$ and $3e$) rules, which are all reversible and equivalence-preserving, we know that our rule is too.
    \end{proof}
    \begin{proof}[Proof (EPR)]
        Trivial, since we retain the information on equivalence classes. But, in more detail, the equivalence preservation of the reductive part of EPR follows from TWSR's proof above, and the equivalence of the substitution part can be proved as follows: Let $x$ and $y$ be in the same strongly connected component of a BIG. Then, we have or can easily deduce the following two clauses: $\overline{x}\lor y$ and $\overline{y}\lor x$. For any clause $C$ containing $y$ we can iterate $\overline{x}\lor y$ within $C$, and deiterate $y$ from it to obtain $C\lor x$. We can then iterate $\overline{y}\lor x$ within the nested area of $y$ (potentially adding a double cut), and deiterate $y$ from it to obtain $y\land x$, which can be deleted by an iteration of the copy of $x$ in expanded $C$, to obtain $C$ with $y$ replaced by $x$.
    \end{proof}
\end{theorem}
Given that all of our rules can never add any variables, literals or clauses, it is also straightforward to prove the following theorem.
\begin{theorem}
    The applications of SWR, EPR, TRR, OSIR, and TWSR are guaranteed to be monotonically non-increasing in the number of variables, in the number of clauses and in the number of literals of a propositional formula.
    \begin{proof}[Proof (EPR)]
        Trivial, since either (i) the BIG has no strongly connected components and so the formula stays the same, or (ii) at least one strongly connected component is found, and for each component, at least the binary clauses corresponding to all the edges in the component can be deleted.
    \end{proof}
    \begin{proof}[Proof (TWSR)]
        Trivial, since either (i) TWSR does not apply so the formula stays the same, (ii) condition (1) is satisfied and so the number of clauses is reduced by one, or (iii) condition (2) applies and the number of literals is reduced by one.
    \end{proof}
\end{theorem}

\section{Algorithmic complexity and future work}
\label{sec:fw}
Our systematic reduction procedure applies TWSR prioritising the propagation of the outermost smallest unprocessed clauses ---since these have greater reduction potential than inner and bigger clauses--- and whenever the smallest unprocessed/new clauses are of size $2$, the procedure applies EPR before propagating them by TWSR. This might seem a straightforward operation, but it becomes a nontrivial task if we want to avoid unnecessarily repeating computations every time a reduction changes the BIG of the formula, or strengthens or modifies an already propagated clause.

\begin{figure}[h]
    \centering
    \includegraphics[width=0.6\textwidth]{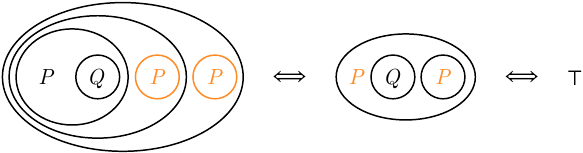}
    \caption{Proof of Peirce’s Law, which states that $((P\rightarrow Q)\rightarrow P)\rightarrow P$, obtained from the systematic application of the EPR\raisebox{0.15ex}{+}TWSR procedure, which prioritises the propagation of the smallest outermost unprocessed clauses. The detected simplifications in each graph transformation are highlighted in orange. Recall that the automatic theorem prover Vampire required 7 steps and 3 flattening transformations (see Fig.~\ref{fig:vampi}).}\vspace{-5pt}
    \label{fig:Plaw}
\end{figure}

Before analysing the complexity of the corresponding algorithm, let us illustrate the potential of our novel approach by revisiting Peirce's Law, which the automatic theorem prover Vampire was able to automatically prove by contradiction in 7 steps, 3 of which were transformations to a normal form (recall Fig.~\ref{fig:vampi}). Note that other state-of-the-art tools, such as the interactive theorem prover Coq\footnote{https://coq.inria.fr/} and MiniZincIDE\footnote{https://www.minizinc.org/} (an IDE to run high-level, solver-independent constraint models), are not able to automatically prove or simplify this tautology. On the contrary, our novel systematic approach can automatically reduce Peirce's Law to $\top$ in as little as 2 steps, as shown in Fig.~\ref{fig:Plaw}.

Note that the double negation (double cut) generated in the first step is automatically simplified by the parser (and is hence not counted as a reasoning step), and the clause deletion whenever an empty cut is generated is part of our TWSR and EPR rules. Nevertheless, even if we count these two trivial simplifications as reasoning steps, our procedure still proves Peirce's Law in half the number of steps required by Vampire, whilst avoiding any form of flattening nor requiring bookkeeping of rule applications (since all of our transformations are equivalence-preserving), and therefore requiring less space too.

In what follows, let $l$ be the total number of literals of a given propositional formula, $v$ the number of variables, and $c$ the total number of clauses across nesting levels, that is, the number of cuts of depth less or equal than $2$ (i.e. containing literals only) not already contained in a cut of depth less or equal than $2$. For example, the middle EG in Figure~\ref{fig:OSIR} has $5$ clauses across nesting levels. Let $s$ be the total number of singletons across all nesting levels, $\beta$ the number of variables present in binary clauses, $\beta'\subseteq\beta$ the number of binary variables present in clauses of size greater than $2$, $\kappa$ the number of binary clauses, $maxL$ the maximal clausal length, and $\mu$ the number of cuts of depth greater than $2$.

Our algorithm stores the given propositional formula as an arborescence of depth $\mu$ corresponding to the EG representation of the formula, the binary clauses in each nesting level as a BIG, and the non-binary clauses as a directed graph with as many nodes as variables and clauses, and as many edges as literals. Even if the BIG might store redundant information due to it needing to be skew-symmetric, and our algorithm computing temporary BIG unions, the size of any temporary helper set used along the computation is minimal compared to the total size of the formula, and less than twofold in the worst case.
Thus, the space complexity of the EPR\raisebox{0.15ex}{+}TWSR procedure is $\mathcal{O}(l)$ in terms of edges, and $\mathcal{O}(v+c)$ in terms of nodes, so \emph{linear} in the size of the formula in any case. Moreover, notice that the size of the working formula will decrease with every EPR\raisebox{0.15ex}{+}TWSR reduction found, so this is clearly its \emph{worst-case space complexity}.

The space complexity of our approach is better than all existing encodings used by PL preprocessors and simplifiers we are aware of, given that, to begin with, we do not need to flatten the original formula before processing. Our approach just transforms arbitrary operators to their equivalent form using $\wedge$ and $\neg$, which has a one-to-one correspondence with the EG representation of the formula. This transformation has no impact on the number of variables, literals or (sub)clauses
. In contrast, rewritings such as the widely used Tseitin transformation~\citep{tseitin1983complexity} need to introduce new variables to avoid potential formula explosions during flattening, and so they not only increase the number of literals and clauses, but might also increase the number of formula variables. The length of the resulting flattened formula might still be linear in the size of the formula, but clearly greater than the one obtained with our approach (recall Example~\ref{ex:sw}).

On the other hand, the \emph{worst-case time complexity} of the EPR\raisebox{0.15ex}{+}TWSR procedure is $\mathcal{O}(l+\mu(\beta\kappa+(\frac{\beta'}{v-\beta+\beta'}\kappa+maxL)(c-\kappa-s)))$, where the term $l$ corresponds to the parsing of the formula, and $\beta\kappa$ encompasses all operations over BIGs, including finding strongly connected components and computing its transitive reduction. The term $(\frac{\beta'}{v-\beta+\beta'}\kappa+maxL)(c-\kappa-s)$ corresponds to the complexity of applying TWSR, where $(c-\kappa-s)$ is the number of clauses of size greater than $2$. Our method avoids unnecessary checks by only comparing clauses which have potential to reduce each other, that is, which share variables or contain variables present in the BIG. The number of binary variables present in each clause of interest is upper bounded by the size of the clause, which is in turn upper bounded by $maxL$. The number of connectivity checks required is very much affected by the proportion of binary variables present in clauses of size greater than $2$, i.e. $\frac{\beta'}{v-\beta+\beta'}$, and the size of the BIG, i.e. $\kappa$. Finally, since EPR and TWSR computations are performed at each nesting level, we need to multiply these complexities by $\mu$.

Notice that if a given formula contains only singletons in its nested cuts, the worst-time complexity collapses to $\mathcal{O}(l)$. If a formula has no binary clauses, then our algorithm's worst time complexity turns into $\mathcal{O}(l+\mu(maxL(c-s))\approx\mathcal{O}(l+\mu c)$, i.e., linear time in the size of the problem. If a given formula has maximum clausal length $2$ ---over all cuts, which note is not the same as a CNF 2-SAT problem--- the complexity is then bounded by $\mathcal{O}(l+\mu\beta\kappa)\approx\mathcal{O}(l+\mu vc)$. 

In summary, the complexity of EPR+TWSR($\varphi$) is linear $\mathcal{O}(c)$ when $s\to c$ and when $\kappa\to 0$. Otherwise, it has quadratic worst-case complexity. In particular, it is upper bounded by $\mathcal{O}(\beta\kappa)$ when $\kappa\to c$, when $\beta'\to 0$ or when $\beta\ll v$. Otherwise, the complexity is upper bounded by $\mathcal{O}(\kappa c)$ and it reaches its maximum $\mathcal{O}(c^2)$ when $\kappa\to(c-s)/2$, that is, when half of the non-singleton clauses are binary. However, sub-quadratic behavior can still be observed in practice on problems with sparse graphs and with non-uniform connectivity distributions ---which is the case for most non-synthetic problems. Moreover, note that the problem size of an unflattened formula can be orders of magnitude smaller than its flattened version, so even if our algorithm is in the same complexity class as other state-of-the-art SAT preprocessing techniques it generalises, its actual complexity can be orders of magnitude smaller.

Another key advantage of our unified approach is that it eliminates the need to choose between different techniques or their order of application, which can have a major impact on the final level of reduction (e.g. some preprocessing techniques add redundant clauses by resolution, whilst others delete them, which could lead to an infinite loop). It can also be the case that some rules are not applicable at first, but they can provide further reduction if applied after other techniques. This is a nontrivial problem usually resolved by heuristics (e.g. restarts) or by limiting the application and behaviour of the rules according to arbitrary parameters (e.g. apply preprocessing to learned or bigger clauses only). Example~\ref{ex:order} illustrates these problems and how our solution addresses them instead. 
Moreover, note that our approach is guaranteed to terminate since it never increases the size of the problem.
\begin{example}\label{ex:order}
    Let $\varphi=(A\lor B\lor C)\land(A\lor B\lor D)\land(B\lor\overline{C})\land(\overline{A}\lor B\lor\overline{C})$. If we apply TWSR to it, our approach first propagates the smallest clause $(B\lor\overline{C})$ and reduces the formula to $\mathrm{TWSR}(\varphi)=\varphi'=(A\lor B)\land(A\lor B\lor D)\land(B\lor\overline{C})$. TWSR next propagates the smallest unprocessed clause $(A\lor B)$, and returns the equivalent formula $\varphi''=(A\lor B)\land(B\lor\overline{C})$. This same result could also be obtained by using other existing techniques combined in many different ways. For example, applying a round of subsumption, then a round of self-subsumption, and then another round of subsumption would achieve the same reduction in this case. We can also obtain $\varphi''$ by applying HTE, HSE and self-subsumption in any order. Adding all the possible redundant clauses obtained by resolution and then applying two rounds of subsumption would also lead to $\varphi''$. Note, however, that the choice of techniques or their application orders would of course be problem-dependent and not obvious beforehand, so many techniques would probably be unsuccessfully applied before reaching (or not) the desired result.
\end{example}

Some of the existing computationally most expensive preprocessing techniques (namely FLP, \textit{hyper binary resolution} (HBR)~\citep{DBLP:conf/aaai/Bacchus02}, \textit{asymmetric subsumption elimination} (ASE)~\citep{DBLP:conf/lpar/HeuleJB10}, \textit{asymmetric literal elimination} (ALE)~\citep{DBLP:series/faia/BiereJK21} and \textit{asymmetric tautology elimination} (ATE)~\citep{DBLP:conf/lpar/HeuleJB10}) can achieve reductions on CNF formulae that TWSR is not able to attain. However, TWSR can reach the same level of reduction if applied to the original nested formula (see Example~\ref{ex:factor}).
\begin{example}\label{ex:factor}
    Let $\varphi=(\overline{C}\lor\neg(\overline{A}\lor\overline{B}\lor\overline{D}))\land(\overline{A}\lor\overline{B}\lor\overline{D})$. Let $\mathrm{CNF}(\varphi)=(\overline{C}\lor A)\land(\overline{C}\lor B)\land(\overline{C}\lor D)\land(\overline{A}\lor\overline{B}\lor\overline{D})$ be $\varphi$ expressed in conjunctive normal form. Both FLP and HBR can reduce $\mathrm{CNF}(\varphi)$ to $\varphi'=\overline{C}\land(\overline{A}\lor\overline{B}\lor\overline{D})$, but TWSR would not be able to do so unless we apply first a nontrivial EG factorisation step to recover its nested form. However, if we apply TWSR directly to $\varphi$, we can obtain $\varphi'$ much more efficiently (see Fig.~\ref{fig:facto}). Note that none of the existing preprocessing methods can be applied directly to $\varphi$ since it is not in CNF.
    \begin{figure}[h]
        \vspace{-5pt}
        \centering
        \includegraphics[width=0.75\textwidth]{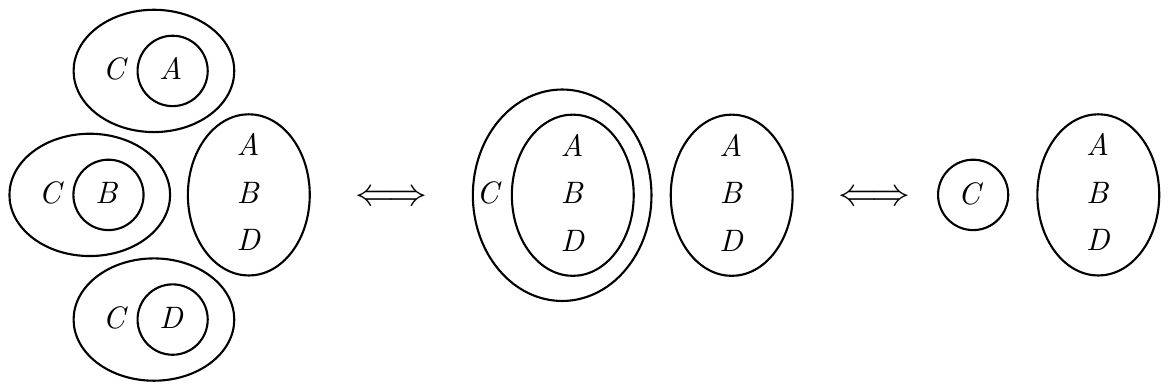}\vspace{-5pt}
        \caption{EGs of $\mathrm{CNF}(\varphi)=(\overline{C}\lor A)\land(\overline{C}\lor B)\land(\overline{C}\lor D)\land(\overline{A}\lor\overline{B}\lor\overline{D})$ (left), its equivalent factorised form $\varphi=(\overline{C}\lor\neg(\overline{A}\lor\overline{B}\lor\overline{D}))\land(\overline{A}\lor\overline{B}\lor\overline{D})$ (middle), and its equivalent reduced form $\mathrm{TWSR}(\varphi)=\varphi'=\overline{C}\land(\overline{A}\lor\overline{B}\lor\overline{D})$ (right).}\vspace{-10pt}
        \label{fig:facto}
    \end{figure}
\end{example}

In fact, by using EGs and BIGs, we have realised that TWSR and these advanced SAT preprocessing techniques can actually be seen as $n$-ary versions of TRR and OSIR, where the nodes of the implication graph can be clauses instead of singletons. That is, instead of finding a redundant edge between two singletons, we remove redundant edges between implied clauses (see Example~\ref{ex:n-tr}), and instead of finding a singleton implying its negation, we uncover an $n$-ary clause implying its negation (see Example~\ref{ex:n-osir}). Thus, we are currently working on an extension of TWSR guided by the \textit{n-ary implication hypergraph} of the formula, which we claim will be able to generalise all the aforementioned rules while still guaranteeing a never-increasing problem size.
\begin{example}\label{ex:n-tr}
    Let $\varphi=(\overline{P}\lor\overline{Q}\lor Y)\land(\overline{Y}\lor X)\land(\overline{X}\lor R\lor S)\land(\overline{P}\lor\overline{Q}\lor R\lor S)$. Note that the first clause can be interpreted as $(P\land Q)\rightarrow Y$, the second as $Y\rightarrow X$, the third as $X\rightarrow(R\lor S)$, and the fourth as $(P\land Q)\rightarrow(R\lor S)$, amongst many other alternative readings. However, with these particular interpretations, we can build the following implication graph:\vspace{-10pt}
    
    \begin{figure}[h]
        \centering
        \vspace{10pt}
        \includegraphics[width=0.35\textwidth]{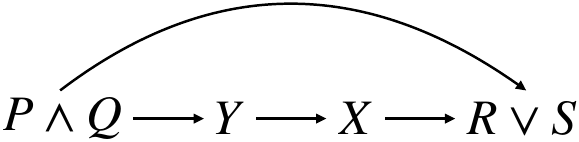}\hspace{2cm}
        \includegraphics[width=0.35\textwidth]{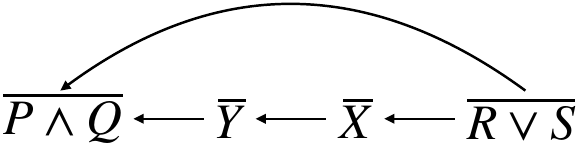}\vspace{-5pt}
    \end{figure}
    
    \noindent from where it is clear that the clause $(\overline{P}\lor\overline{Q}\lor R\lor S)$ corresponds to a redundant edge and so it can be deleted. This reduction can therefore be seen as an $n$-ary version of TRR. Our current version of TWSR cannot find these redundancies, nor can FLP or HBR. From the rules we have mentioned in this paper, only ASE and ATE would be able to detect this redundancy.
\end{example}
\begin{example}\label{ex:n-osir}
    Let $\varphi=(\overline{X}\lor Y)\land(\overline{Y}\lor\overline{Z})\land(X\lor\overline{P}\lor\overline{Q})\land(\overline{Y}\lor\overline{P}\lor\overline{Q})$. Remember that we can read $(\overline{X}\lor Y)$ as $(X\rightarrow Y)$ or, equivalently, $(\overline{Y}\rightarrow\overline{X})$. The clause $(X\lor\overline{P}\lor\overline{Q})$ can be interpreted as $(P\land Q\rightarrow X)$ or $(\overline{X}\rightarrow\overline{(P\land Q)})$, amongst many other equivalent readings. Hence, the reduction obtained from the application of TWSR, namely $\varphi'=\mathrm{TWSR}(\varphi)=(\overline{X}\lor Y)\land(\overline{Y}\lor\overline{Z})\land(\overline{P}\lor\overline{Q})$, can also be seen as applying an $n$-ary version of OSIR to the following implication path: $P\land Q\implies\overline{Y}\implies\overline{X}\implies\overline{(P\land Q)}$.
\end{example}

\section{Conclusion}
\label{sec:conc}
As mentioned earlier, reasoning with EGs allowed us to independently rediscover many known equivalence-preserving SAT preprocessing techniques, gain a better understanding of their underlying relationships, as well as prove and establish more directly many of their properties. EGs do not only offer us a fresh  viewpoint on logical reasoning, but helped us to build an entirely different and novel simplification approach which generalises many of these techniques with added advantages: it is more efficient, avoids look-ahead backtracking, guarantees a monotonically decreasing number of variables, literals and clauses (and so termination), it is structure-preserving, can be applied to nested formulae, and does not require a propositional formula to be given in CNF.

With our approach, it becomes clear why some simplification techniques, based on adding redundant clauses or literals, can sometimes help to reduce a problem but other times do not, and may instead lead to wasted preprocessing effort and the need for bespoke or contrived heuristics. Since our proposed method generalises a significant set of previously-thought independent techniques, the high complexity and effort associated with finding a suitable application order are drastically reduced.
As described in detail in the previous section, our rules can also decrease the space complexity of the problem since they can be applied before formulas are flattened (e.g. converted to CNF). This can greatly minimise the search space and even prevent potential explosions of the formula size during translation.
Our reductions also contribute to a deeper understanding of the problem at hand, which leads to better modelling, solving strategies, search heuristics and translations, as well as further insights into breaking symmetry, providing model counting bounds and aiding \#SAT. Moreover, our techniques are solver-, problem-, and (normal) form-agnostic and apply to propositional logic formulae in general. Hence, we expect them to be equally beneficial in other fields such as SMT, automated reasoning and theorem proving.

Finally, directions for future work include extensions of the TWSR rule informed by a novel $n$-ary implication hypergraph, refining the current implementation of our algorithm, and a broader experimental evaluation.

%
%
\bibliographystyle{tlplike}
\bibliography{mybib}

\end{document}